# Search for slowly moving magnetic monopoles

J. T. Hong
Department of Physics, Boston University, Boston, Massachusetts 02215, U.S.A.*
and California Institute of Technology, Pasadena, California 91125, U.S.A.
(For the MACRO Collaboration)

We report a search for slowly moving magnetic monopoles in the cosmic radiation with the first supermodule of the MACRO detector at Gran Sasso. The absence of candidates established an upper limit on the monopole flux of $5.6 \times 10^{-15}$ cm$^{-2}$sr$^{-1}$s$^{-1}$ at 90% confidence level for the velocity range of $1.8 \times 10^{-4} < \beta < 3 \times 10^{-3}$.

Supermassive magnetic monopoles are predicted in Grand Unified Theories (GUTs) [1]. Produced in the early universe, relic monopoles in the galaxy are expected to travel with typical galactic velocities of $\sim 10^{-3}c$ and their flux is expected not to exceed the astrophysical Parker bound [2]. It has been argued [3,4] that monopoles may be trapped in the solar system and thus their local flux may be enhanced above the Parker bound. We concentrate in this paper on a search for these trapped monopoles that are expected to travel at velocities as low as $\sim 10^{-4}c$ (the orbital velocity about the Sun at 1 A.U.).

The MACRO detector [5], a large underground detector, is nearing completion at the Gran Sasso Laboratory in Italy, with the primary goal of searching for monopoles at flux levels below the Parker bound. The experimental data reported in this paper uses only the first operational supermodule of the MACRO detector, which is documented in detail elsewhere [5,6]. Its dimensions are $12.6 \times 12 \times 4.8$ m$^3$ and its acceptance is 870 m$^2$sr. It is surrounded on five sides by planes of liquid scintillator counters: 32 horizontal counters in two horizontal planes and 21 vertical counters in three vertical planes. Each counter is an 11 m long tank of liquid scintillator viewed by 20 cm diameter hemispherical photomultiplier tubes. A horizontal counter has two phototubes at each end, while a vertical counter has only one phototube at each end. In addition, there are limited streamer tubes, plastic track-etch detectors, and passive rock absorber.

In the liquid scintillator subsystem, we have employed two types of specialized monopole triggers, which cover different $\beta$ regions [6]. In this paper, we report the monopole search results from data collected using the trigger which covers the lower $\beta$ region [6] and has also been applied to a search for nuclearites (strange quark matter) [7]. Based on the time of passage of slowly moving particles through the 19 cm thick scintillator, this trigger selects wide phototube pulses or long trains of single photoelectron pulses generated by slowly moving particles, and rejects with high efficiency any single large but narrow pulse from the residual penetrating cosmic ray muons or from radioactive decay products from the environment. The sensitivity of this trigger to slow monopoles was measured by simulating the expected signals using light-emitting diodes (LEDs) in representative counters [6,8]. To discriminate monopole signals against backgrounds, waveforms of the phototube signal are recorded by two complementary sets of waveform digitizers.

The data were collected from October 1989 to November 1991 with an accumulated live time of 542 days, during which there were 583,999 events with the slow monopole trigger present in at least one scintillator plane. After vetoing events which also fired a two-plane muon trigger requiring time-of-flight $< 1 \mu$s (corresponding to $\beta > 1.5 \times 10^{-2}$), 541,918 events remained. The majority of these single plane monopole triggers were due to radioactivity pileups (*i.e.*, many background radioactivity-induced pulses accidentally occurring within a short time interval). Requiring the trigger to be present in two separate planes within 600 $\mu$s (the time-of-flight for a $\beta = 10^{-4}$ particle to cross the apparatus with the longest possible path length), as expected for slow monopoles, yielded 723 events, some of which were caused by power glitches which eventually ended the run. To eliminate these, candidate events were then required to occur at least 0.015 hour before the end-of-run. This end-of-run cut reduced the total live time to 541 days and 573 candidates survived, each of which was examined and classified using a waveform analysis.

---

*Current address.

The majority of these candidates (565 events) were easily identified as due to electrical noise by the following: the presence of bipolar oscillations in their recorded waveforms (388); having no feature other than occasional isolated radioactivity-induced pulses in the waveform (169), interpreted as being caused by electrical noise on the trigger input; or having long pulse trains ($> 4\,\mu s$) simultaneously present in every channel (8), inconsistent with passage of particles.

The remaining eight candidates are of non-electrical origins. Among them, two candidates were identified as muons because of their time-of-flight and pulse shapes; they escaped the fast muon veto because they occurred during a period when the fast muon trigger malfunctioned. Three other candidates had muon signals in one plane and radioactivity pileups in the other plane.

The remaining three candidates had waveforms consisting of 4–8 narrow pulses in sequence, where each pulse typically had a pulse height at least several times larger than the average single photoelectron pulse height, and therefore inconsistent with the expectation for monopoles. Instead, these events are consistent with being due to accidental coincidences between radioactivity pileups in different planes, for which the expected number is calculated to be 2.6, compared to the three that were observed. We note that for the passage of slow particles, the photoelectrons should be randomly but uniformly distributed to produce much smoother pulse trains than the observed "spiky" pulse trains. A quantitative analysis of the "spikiness" of a pulse train has rejected these three candidates [8]. The candidate with the least spiky waveforms is shown in Fig. 1. In addition to being too spiky, the pulse train durations are inconsistent with the time-of-flight unless we consider

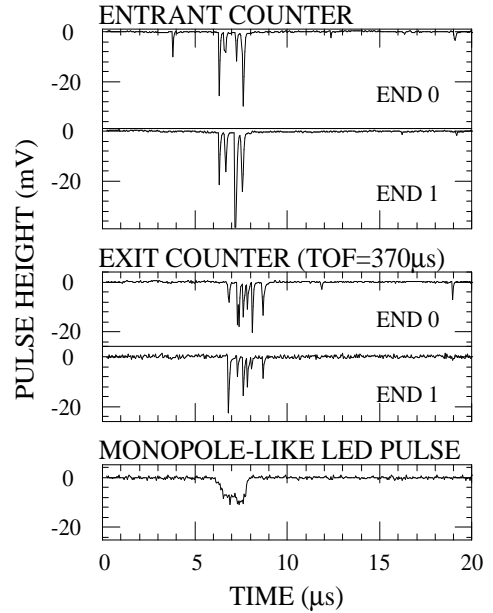

Figure 1. The candidate with the least spiky waveforms compared with an LED-generated monopole-like pulse of the approximately same duration and integrated area.

the trajectories clipping corners of the scintillator counters.

As a final cross-check, we have looked at the streamer tubes for slow particle triggers or tracks for the sample of the 573 candidates, and found agreement with the above classifications based on the scintillator waveforms. For future more sensitive searches, the streamer tube subsystem will give an additional strong handle. Furthermore, we note that if any candidate events survive, we can perform a rigorous inspection of the tracks in the track-etch subsystem.

In conclusion, we have found no evidence for the passage of a slowly moving ionizing particle through MACRO and have established upper limits on the fluxes of GUT monopoles and dyons shown in Fig. 2 [9–13], assuming the flux is isotropic when the particle encounters the Earth. As indicated by the bold solid curve, the flux limit for bare monopoles is $5.6 \times 10^{-15} \mathrm{cm}^{-2}\mathrm{sr}^{-1}\mathrm{s}^{-1}$

(90% confidence level) in the velocity range of $1.8 \times 10^{-4} < \beta < 3 \times 10^{-3}$. An additional low-$\beta$ range of $1.5 \times 10^{-4} < \beta < 1.8 \times 10^{-4}$ with the flux limit of $1.3 \times 10^{-14} \text{cm}^{-2}\text{sr}^{-1}\text{s}^{-1}$ is established using exclusively the more sensitive horizontal counters. Due to trajectories with path lengths through scintillator longer than the minimum 19 cm, the upper $\beta$ limit extends up to $4 \times 10^{-3}c$, but with a less restrictive flux limit. Because of their larger nuclear stopping power, dyons or monopole-proton composites with velocity $\beta \sim 10^{-4}$ can not penetrate the Earth. Therefore, the dyon flux limit at the lower part of the $\beta$ range is modified from the bare monopole flux limit using the composition and density profile of the Earth and the results are indicated as dashed curves in Fig. 2 extending down to $8 \times 10^{-5}c$. Bracci et al. [14] have argued that bare monopoles are likely to have captured and bound protons in the early universe, making this dyon search especially relevant.

Finally, since an encounter with a star is the proposed mechanism for slowing down and trapping a monopole in the solar system [3,4] and since a monopole emerging from such an encounter will have attached a proton, we emphasize that the flux limits presented here place new constraints on the abundance of trapped monopoles.

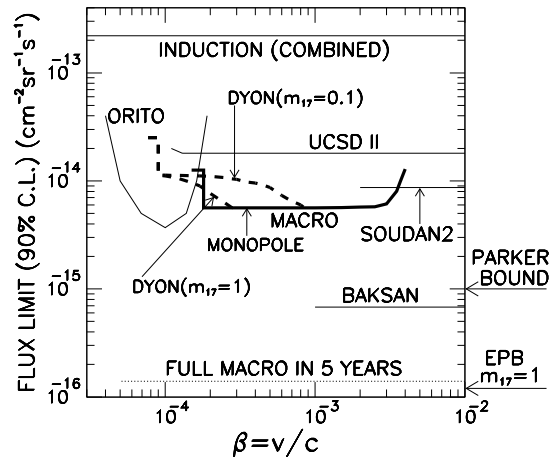

Figure 2. The upper limits on the fluxes of bare monopoles and dyons with mass $10^{17}\,\text{GeV}/c^2$ ($m_{17} = 1$) or $10^{16}\,\text{GeV}/c^2$ ($m_{17} = 0.1$) at 90% confidence level as a function of the velocity with which a monopole or dyon enters the Earth. Also shown are the anticipated limit for bare monopoles reachable by the full MACRO detector after five years of operation, the Parker bound[2], the extended Parker bound (EPB) for monopoles of mass $10^{17}\,\text{GeV}/c^2$ ($m_{17} = 1$) [15], and the results for bare monopoles from several previous searches: Induction (combined)[9], UCSD II (He-$CH_4$) [10], Soudan 2 (Ar-$CO_2$) [11], Baksan (scintillator) [12], and Orito (CR-39) [13].